\documentstyle[11pt,epsf]{article}
 1
 1
 1

\def\lsim{\mathrel{\raise.3ex\hbox{$<$\kern-.75em\lower1ex\hbox{$\sim$}}}}
\def\gsim{\mathrel{\raise.3ex\hbox{$>$\kern-.75em\lower1ex\hbox{$\sim$}}}}
\catcode`\@=11
\def\slash{\mathpalette\make@slash}
\def\make@slash#1#2{\setbox\z@\hbox{$#1#2$}%
  \hbox to 0pt{\hss$#1/$\hss\kern-\wd0}\box0}
\catcode`\@=12 
\begin{document}
\noindent
\thispagestyle{empty}
\renewcommand{\thefootnote}{\fnsymbol{footnote}}
\begin{flushright}
{\bf UCSD/PTH 97-38}\\
{\bf hep-ph/9801273}\\
{\bf January 1998}\\
\end{flushright}
\vspace{.5cm}
\begin{center}
  \begin{Large}\bf
Top Quark Pair Production at Threshold -- \\[1mm]
Uncertainties and Relativistic Corrections
\footnote{Invited talk presented at the Workshop on Physics at the
First Muon Collider and the Front End of a Muon Collider, Fermilab,
November 6 - 9, 1997, to appear in the proceedings.}
%
  \end{Large}
  \vspace{1.5cm}

\begin{large}
 A.H. Hoang
\end{large}
\begin{center}
\begin{it}
   Department of Physics,
   University of California, San Diego,\\
   La Jolla, CA 92093--0319, USA\\ 
\end{it} 
\end{center}

  \vspace{1.5cm}
  {\bf Abstract}\\
\vspace{0.3cm}
%
\noindent
\begin{minipage}{15.0cm}
\begin{small}
In this talk it is shown how nonrelativistic QCD (NRQCD) can be used
to determine next-to-next-to-leading order relativistic and
short-distance contributions to the total $t\bar t$ production cross
section in the threshold
regime at lepton colliders. A recipe for the calculation of all such
contributions for the total photon mediated production cross section
is presented and a review of the already known Abelian
next-to-next-to-leading results is given.
\end{small}
\end{minipage}
\end{center}
\setcounter{footnote}{0}
\renewcommand{\thefootnote}{\arabic{footnote}}
\vspace{1.2cm}

\section*{Introduction}
The production of $t\bar t$ pairs in the threshold region at future
lepton colliders like the NLC (Next Linear Collider) or the FMC (First
Muon Collider) offers a unique opportunity to carry out precision
tests of QCD in a completely new environment. Due to the large top
mass\footnote{
Throughout this talk $M_t$ is understood as the top quark pole mass.
}
($M_t\approx 175$~GeV), which allows for the decay channel $t\to W b$,
hadronization effects can be neglected in a first
approximation~\cite{Fadin1}.
This makes the $t\bar t$ production cross section in the threshold
regime (including various distributions) calculable from perturbative 
QCD (and electroweak interactions), which then allows for precise
extractions of the top quark mass and the strong coupling once the
cross section is measured. In fact, experimental simulations for the
NLC and the FMC~\cite{NLC,Berger1} have shown that experimental errors of
around $100$~MeV for the top quark mass and of around $0.002$ for
$\alpha_s(M_z)$ can be expected for a cross section measurement with a
total integrated luminosity of $50-100 fb^{-1}$. In particular the
prospect for the error in the top quark mass measurement beats any
hadron collider experiment. However, the errors for $M_t$ and
$\alpha_s$ given above do not contain any theoretical
uncertainties. At this point it is illustrative to recall that the
standard present day formalism used for describing $t\bar t$
production at threshold consists of solving the nonrelativistic
Schr\"odinger equation with a QCD potential which for small distances
is given by perturbative QCD up to one loop~\cite{Fischler1,Billoire1}
and for large 
and intermediate distances by fits to quarkonia spectra (and leptonic
decay widths)~\cite{Strassler1,QCDpotential}. The results are then
modified by various
${\cal{O}}(\alpha_s)$ short-distance corrections which makes the
results correct at the next-to-leading order (NLO) level, i.e., they
properly include all ${\cal{O}}(\alpha_s)$ corrections.\footnote{
The solutions of the nonrelativistic Schr\"odinger equation with the
one-loop corrected QCD potential contains, in the language of Feynman
diagrams, the resummation of terms $\propto
(\alpha_s/v)^n\times [1,\alpha_s]$, $n=0,1,\ldots,\infty$, ($v$
being the top quark c.m.\ velocity) to all orders in
$\alpha_s$. Because in the threshold region $|v|\lsim\alpha_s$ we
count all terms $\propto \alpha_s/v$ of order one.
}
NNLO (${\cal{O}}(\alpha_s^2)$) corrections have never been taken
into account so far. Their contributions, however, can be sizable. As
an example, consider the ${\cal{O}}(\alpha_s^2)$ relativistic
corrections to the total cross sections which can lead to a shift in
the 
location of the 1S peak of order $M_t\alpha_s^4\sim 150$~MeV and a
corrections of order $\alpha_s^2\sim 3\%$ in the size of the cross
section (for $\alpha_s\sim\alpha_s(M_t\alpha_s)\sim 0.17$). Further,
even the  
${\cal{O}}(\alpha_s^2)$ short-distance corrections normalizing the
total cross sections might be large if the huge size of the
${\cal{O}}(\alpha_s)$ corrections of order $-20\%$ is taken into
account. From this point of view it is clear that the theoretical
uncertainties in the present day analyses are certainly not negligible
and that full control over all NNLO effects should be gained. 

Unfortunately the formalism described above is constructed in a way
that makes a systematic and rigorous implementation of all NNLO
effects 
from first principles QCD conceptually difficult if not impossible --
a consequence of the use of phenomenological information in the
potential for large and intermediate distances. In principle, this
formalism has to be considered as a (sophisticated) potential model
approach which cannot be improved in a rigorous way at all. I
therefore propose to 
rely on
perturbative QCD only. This means that one applies the perturbative
QCD potential for all distances. In fact, such a decision seems to be
just natural if one takes
into account that the $t\bar t$ system is almost insensitive to the
large distance (i.e.\ non-perturbative) contributions in the QCD
potential. This makes the framework in which effects beyond the NLO
level shall be determined more transparent and still leaves the
possibility to incorporate the non-perturbative effects later as a
perturbation (see e.g.~\cite{Yakovlev1} for an approach of this sort).

In this talk I demonstrate how NRQCD~\cite{Caswell1} can be used
to determine NNLO relativistic and short-distance corrections to the
total $t\bar t$ production cross section. For simplicity only 
the production through a virtual photon is
considered. The generalization for production through different
currents is straightforward. I want to stress that I do not
talk about the 
peculiar NNLO finite width effects coming from the
off-shellness of the top quark, time dilatation effects and the
interaction among the decay products. These effects have been addressed
previously 
in a number of publications~\cite{Sumino1,Teubner1,Moedritsch1}, but
no rigorous and consistent
description of them for the $t\bar t$ cross section has been found
yet. To achieve full NNLO accuracy these finite width effects should
eventually be included. For now they remain as an unsolved
problem. As far as the NNLO relativistic corrections discussed in this
talk are concerned I will use the naive replacement
\begin{equation}
E\equiv\sqrt{s} - 2 M_t \longrightarrow \tilde E = E + i \, \Gamma_t
\label{replacement}
\end{equation}
in the spirit of~\cite{Fadin1} in order to examine their size and
properties, where $\Gamma_t$ represents a constant which
is not necessarily the decay width of a free top quark.

\section*{The \lowercase{$t\bar t$} Cross Section in NRQCD}
NRQCD is an effective field theory of QCD designed to handle
nonrelativistic systems of heavy quark--antiquark pairs to in principle
arbitrary precision. It is based on the separation of long- and
short-distance effects\footnote{
In this context ``long-distance'' is not equivalent to
``non-perturbative''.
}
by reformulating QCD in terms of a non-renormalizable Lagrangian
containing all possible operators in accordance to the symmetries in
the nonrelativistic limit. The NRQCD Lagrangian reads
\begin{eqnarray}
{\cal{L}}_{\mbox{\tiny NRQCD}} & = &
- \frac{1}{2} \, \mbox{Tr} \,G^{\mu\nu} G_{\mu\nu} 
+ \sum_{q=u,d,s,c,b} \bar q \, i \slash{D} \, q
\nonumber\\[2mm] & &
+\, \psi^\dagger\,\bigg[\,
i D_t 
+ a_1\,\frac{{\mbox{\boldmath $D$}}^2}{2\,M_t} 
+ a_2\,\frac{{\mbox{\boldmath $D$}}^4}{8\,M_t^3}
+ 
\frac{a_3\,g}{2\,M_t}\,{\mbox{\boldmath $\sigma$}}\cdot
    {\mbox{\boldmath $B$}}
\nonumber\\[2mm] & & \quad
+ \, \frac{a_4\,g}{8\,M_t^2}\,(\,{\mbox{\boldmath $D$}}\cdot 
  {\mbox{\boldmath $E$}}-{\mbox{\boldmath $E$}}\cdot 
  {\mbox{\boldmath $D$}}\,)
+ \frac{a_5\,g}{8\,M_t^2}\,i\,{\mbox{\boldmath $\sigma$}}\,
  (\,{\mbox{\boldmath $D$}}\times 
  {\mbox{\boldmath $E$}}-{\mbox{\boldmath $E$}}\times 
  {\mbox{\boldmath $D$}}\,)
 \,\bigg]\,\psi 
+\ldots
\,.
\label{NRQCDLagrangian}
\end{eqnarray}
The gluonic and light quark degrees of freedom are described by the
conventional relativistic Lagrangian, whereas the top and antitop
quarks are described by the Pauli spinors $\psi$ and $\chi$,
respectively. For convenience all color indices are suppressed. The
straightforward antitop bilinears are omitted and
only those terms relevant for the NNLO cross section are displayed.
$D_t$ and {\boldmath $D$} are the time and space components of the
gauge covariant derivative $D_\mu$ and $E^i = G^{0 i}$ and $B^i =
\frac{1}{2}\epsilon^{i j k} G^{j k}$ the electric and magnetic
components of the gluon field strength tensor (in Coulomb gauge).
The short-distance coefficients $a_1,\ldots,a_5$ are normalized to one
at the Born level. 

To formulate the normalized total $t\bar t$ production cross
section (via a virtual photon) 
$R = \sigma({e^+e^- \atop \mu^+\mu^-}\to\gamma^*\to t\bar
t)/\sigma_{pt}$ 
($\sigma_{pt} =4\pi\alpha^2/3s$) in the nonrelativistic region at NNLO
in NRQCD we start from the fully covariant expression for the cross
section
\begin{equation}
R(q^2) \, = \,
\frac{4\,\pi\,Q_t^2}{s}\,\mbox{Im}[\,
\langle\, 0\,|\, 
T\, \tilde j_\mu(q) \,\tilde j^\mu(-q)\, |\,0\,\rangle]
\,,
\label{crosssectioncovariant}
\end{equation}
where $Q_t=2/3$ is the electric charge of the top quark.
We then expand the electromagnetic current (in momentum space)
$\tilde j_\mu(\pm q) = (\tilde{\bar t}\gamma^\mu \tilde t)(\pm q)$
which produces/annihilates
a $t\bar t$ pair with c.m.\ energy $\sqrt{q^2}$ in terms of ${}^3\!S_1$
NRQCD currents up to dimension eight ($i=1,2,3$)\footnote{
Only the spatial components of the current contribute. 
The expansion of $\tilde j_\mu(-q)$ is obtained from
Eq.~(\ref{currentexpansion}) via charge conjugation
symmetry.
}
\begin{equation}
\tilde j_i(q) = b_1\,\Big({\tilde \psi}^\dagger \sigma_i 
\tilde \chi\Big)(q) -
\frac{b_2}{6 M_t^2}\,\Big({\tilde \psi}^\dagger \sigma_i
(\mbox{$-\frac{i}{2}$} 
\stackrel{\leftrightarrow}{\mbox{\boldmath $D$}})^2
 \tilde \chi\Big)(q) + \ldots
\,,
\label{currentexpansion}
\end{equation}
where the constants $b_1$ and $b_2$ are short-distance coefficients
normalized to one at the Born level. 
Inserting expansion~(\ref{currentexpansion}) back into
Eq.~(\ref{crosssectioncovariant}) leads to the NRQCD expression
of the nonrelativistic cross section at the NNLO level
\begin{eqnarray}
R_{\mbox{\tiny NNLO}}^{\mbox{\tiny thr}}(\tilde E) & = &
\frac{\pi\,Q_t^2}{M_t^2}\,C_1(\mu_{\rm hard},\mu_{\rm fac})\,
\mbox{Im}\Big[\,
{\cal{A}}_1(\tilde E,\mu_{\rm soft},\mu_{\rm fac})
\,\Big]
\nonumber\\[2mm]
& & - \,\frac{4 \, \pi\,Q_t^2}{3 M_t^4}\,
C_2(\mu_{\rm hard},\mu_{\rm fac})\,
\mbox{Im}\Big[\,
{\cal{A}}_2(\tilde E,\mu_{\rm soft},\mu_{\rm fac})
\,\Big]
+ \ldots
\,,
\label{crosssectionexpansion}
\end{eqnarray}
where
\begin{eqnarray}
{\cal{A}}_1 & = & \langle \, 0 \, | 
\, ({\tilde\psi}^\dagger \vec\sigma \, \tilde \chi)\,
\, ({\tilde\chi}^\dagger \vec\sigma \, \tilde \psi)\,
| \, 0 \, \rangle
\,,
\label{A1def}
\\[2mm]
{\cal{A}}_2 & = & \mbox{$\frac{1}{2}$}\,\langle \, 0 \, | 
\, ({\tilde\psi}^\dagger \vec\sigma \, \tilde \chi)\,
\, ({\tilde\chi}^\dagger \vec\sigma \, (\mbox{$-\frac{i}{2}$} 
\stackrel{\leftrightarrow}{\mbox{\boldmath $D$}})^2 \tilde \psi)\,
+ \mbox{h.c.}\,
| \, 0 \, \rangle
\,.
\label{A2def}
\end{eqnarray}
The cross section is expanded in terms of a sum of absorptive parts of
nonrelativistic current-current correlators (containing long-distance
physics) multiplied by short-distance coefficients $C_i$
($i=1,2,\ldots$). In Eq.~(\ref{crosssectionexpansion})
I have also shown the dependences on the various renormalization
scales: the soft scale $\mu_{\rm soft}$ and the hard scale $\mu_{\rm hard}$ are
governing the perturbative expansions of the correlators and the
short-distance coefficients\footnote{
The scales $\mu_{\rm soft}$ and $\mu_{\rm hard}$ arise from the light degrees
of freedom in ${\cal{L}}_{\mbox{\tiny NRQCD}}$ and are already present in
full QCD, whereas $\mu_{\rm fac}$ is generated by ``new'' UV divergences
in NRQCD diagrams. It is crucial to consider $\mu_{\rm soft}$ and
$\mu_{\rm hard}$ as independent scales. Because both, nonrelativistic
correlators and short-distance coefficients, are calculated
perturbatively a residual dependence on $\mu_{\rm soft}$ and $\mu_{\rm hard}$
remains. 
}, 
respectively, whereas the factorization scale $\mu_{\rm fac}$ essentially
represents the boundary between hard (i.e.\ of order $M_t$) and soft
momenta. Because this boundary can in principle be chosen freely,
both, 
correlators and the short-distance coefficients, in general depend on
it (leading to new anomalous dimensions). Because the term in the
second line in Eq.~(\ref{crosssectionexpansion}) is already of NNLO
(i.e.\ suppressed by $v^2$) we can set $C_2=1$ and ignore the
factorization scale dependence of the correlator ${\cal{A}}_2$.
The calculation of all terms in
expression~(\ref{crosssectionexpansion}) proceeds in two basic steps.
\begin{itemize}
\item[1.]{\it Calculation of the nonrelativistic correlators.} --
Determination of the correlators in Eq.~(\ref{crosssectionexpansion})
by taking into account the interactions up to NNLO displayed in
${\cal{L}}_{\mbox{\tiny NRQCD}}$.
\item[2.]{\it Matching calculation.} -- Calculation of the constant
$C_1$ up to ${\cal{O}}(\alpha_s^2)$ by matching
expression~(\ref{crosssectionexpansion}) to the fully covariant cross
section at the two-loop level in the (formal) limit
$\alpha_s\ll v\ll 1$ where an expansion in (first) $\alpha_s$ and
(then) $v$ is feasible. 
\end{itemize} 
{\it Calculation of the nonrelativistic correlators:} In Coulomb gauge
the gluon propagation is separated into a {\it longitudinal},
instantaneous (i.e.\ energy-independent) and a {\it transverse},
non-instantaneous (i.e.\ energy-dependent) propagation. The
longitudinal exchange between the $t\bar t$ pair is described by an
instantaneous potential. (The Coulomb potential is just the LO
interaction caused by the longitudinal exchange.) The transverse
exchange, however, leads to a temporally retarded interaction,
closely related to Lamb-shift type effects known in
QED. Fortunately, because the $t\bar t$ pair is produced/annihilated
in a color singlet (S-wave) configuration, the energy dependence
of the transverse gluon exchange 
leads only to NNNLO (i.e.\ ${\cal{O}}(\alpha_s^3)$ relative to the
effects of the LO Coulomb exchange) contributions.\footnote{
This can be seen by either using formal counting rules 
(see e.g.~\cite{Labelle1,Grinstein1})
or from positronium results where this phenomenon is well
known. From the physical point of view the suppression comes from the
fact that the transverse gluon is radiated \underline{after} the
$t\bar t$ pair is produced and absorbed \underline{before} the $t\bar
t$ pair is annihilated. This process is already suppressed by
$v^2$ due to the dipole matrix element for transverse gluon
radiation/absorption. If the gluon also carries energy, another (phase
space) factor $v$ arises because the gluon essentially becomes
real. For the same reason no top quark self energy or crossed ladder
diagrams have to be considered.
}
We therefore ignore the energy dependence of the transverse gluons which
allows us to formulate \underline{all} interactions contained in the
NRQCD Lagrangian in terms of instantaneous potentials. In other words,
as far as the nonrelativistic correlators at NNLO in
Eq.~(\ref{crosssectionexpansion}) are concerned, NRQCD reduces to a
two-body (top-antitop) Schr\"odinger theory. The potential in the
resulting Schr\"odinger equation is determined by considering $t\bar
t\to t\bar t$ one gluon exchange t-channel scattering amplitudes in
NRQCD. To NNLO 
(i.e.\ including potentials suppressed by at most $\alpha_s^2$,
$\alpha_s/M_t$ or 
$1/M_t^2$ relative to the Coulomb potential) all potentials are
already known and read ($a\equiv C_F\,\alpha_s(\mu_{\rm soft})$, 
$C_F=4/3$, $C_A=N_c=3$, $r\equiv | \vec r |$)
\begin{eqnarray}
V_{\mbox{\tiny BF}}(\vec r) & = & 
\frac{a\,\pi}{M_t^2}\,
\Big[\,
1 + \frac{8}{3}\,\vec S_t \,\vec S_{\bar t}
\,\Big]
\,\delta^{(3)}(\vec r)
+ \frac{a}{2\,M_t^2\,r}\,\Big[\,
\vec\nabla^2 + \frac{1}{r^2} \vec r\, (\vec r \, \vec\nabla) \vec\nabla
\,\Big]
\nonumber\\[2mm] & &
- \,\frac{3\,a}{M_t^2\,r^3}\,
\Big[\,
\frac{1}{3}\,\vec S_t \,\vec S_{\bar t} - 
\frac{1}{r^2}\,\Big(\vec S_t\,\vec r\,\Big)
\,\Big(\vec S_{\bar t}\,\vec r\,\Big)
\,\Big]
+ \frac{3\,a}{2\,M_t^2\,r^3}\,\vec L\,(\vec S_t+\vec S_{\bar t})
\label{BFpotential}
\,,\\[2mm]
V_{\mbox{\tiny NA}}(\vec r) & = & -\,\frac{C_A}{C_F}\,
\frac{a^2}{2\,M_t\,r^2}
\,,
\label{NApotential}
\end{eqnarray}
where $\vec S_t$ and $\vec S_{\bar t}$ are the top and antitop spin
operators and $\vec L$ is the angular momentum operator.
$V_{\mbox{\tiny BF}}$ is the Breit-Fermi potential known from
positronium and $V_{\mbox{\tiny NA}}$ a purely non-Abelian potential generated
through non-analytic terms in one-loop NRQCD (or QCD) diagrams
containing the triple gluon vertex (see
e.g.~\cite{nonabelianpotential} for an older 
reference). The Coulomb potential $V_{c}(\vec
r)=-a/r\,[1+\mbox{corrections up to
${\cal{O}}(\alpha_s^2)$}]$  is not displayed due to lack of space. Its
${\cal{O}}(\alpha_s)$ and ${\cal{O}}(\alpha_s^2)$ corrections have
been determined in~\cite{Fischler1,Billoire1} and \cite{Peter1},
respectively. 

The nonrelativistic correlators are directly related to the Green
function of the Schr\"odinger equation
\begin{eqnarray}
\bigg(\,-\frac{\vec\nabla^2}{M_t} - \frac{\vec\nabla^4}{4M_t^3} +
V_{c}(\vec r) + V_{\mbox{\tiny BF}}(\vec r) + V_{\mbox{\tiny
NA}}(\vec r)  
-\tilde E\,\bigg)\,G(\vec r,\vec r^\prime,\tilde E)
& = & \delta^{(3)}(\vec r-\vec r^\prime) 
\,,
\label{Schroedingerfull}
\end{eqnarray}
where $V_{\mbox{\tiny BF}}$ is evaluated for the ${}^3\!S_1$ configuration
only. The correlators read
\begin{eqnarray}
{\cal{A}}_1 & = & 6\,N_c\,\Big[\,
\lim_{|\vec r|,|\vec r^\prime|\to 0}\,G(\vec r,\vec r^\prime,\tilde E)
\,\Big]
\,,
\label{A1Greenfunctionrelation}
\\[2mm]
{\cal{A}}_2  & = & M_t\,\tilde E\,{\cal{A}}_1
\,.
\label{A2Greenfunctionrelation}
\end{eqnarray}
Relation~(\ref{A1Greenfunctionrelation}) can be easily inferred by
taking into account that the Green function $G(\vec r,\vec
r^\prime,\tilde E)$ describes the propagation of a top-antitop pair
which is produced and annihilated at distances $|\vec r|$ and $|\vec
r^\prime|$, respectively. (A more formal derivation can be found
e.g. in~\cite{Strassler1}.) Relation~(\ref{A2Greenfunctionrelation}),
on the other hand, is obtained through the equation of
motion~(\ref{Schroedingerfull}). Because the exact solution of
Eq.~(\ref{Schroedingerfull}) seems to be an impossible task and we 
start with the well-known Coulomb Green function
$G_c^{(0)}$~\cite{Wichmann1}
($V_c^{(0)}(\vec r)\equiv -a/r$),
\begin{equation}
\bigg(\,-\frac{\nabla^2}{M_t} - V_c^{(0)}(\vec r)
-\tilde E\,\bigg)\,G_c(\vec r,\vec r^\prime,\tilde E)
\, = \, \delta^{(3)}(\vec r-\vec r^\prime)\,,
\label{SchroedingerNR}
\end{equation}
and incorporate the higher order terms using Rayleigh-Schr\"odinger
time-independent perturbation theory (TIPT). It should be noted that
the limit $|\vec r|,|\vec r^\prime|\to 0$ causes UV divergences which
have to be 
regularized using either an explicit short-distance cutoff or
dimensional 
regularization. The freedom in the choice of the regularization
parameter is the origin of the factorization scale 
dependence mentioned before.\\[2mm]
{\it Matching calculation:}
The determination of the short-distance coefficient $C_1$ up to
${\cal{O}}(\alpha_s^2)$ can be carried out in two ways: One either
calculates the constants $b_1$ in Eq.~(\ref{currentexpansion}) through
two-loop matching at the amplitude level for the electromagnetic
vertex in full QCD and NRQCD or one matches
expression~(\ref{crosssectionexpansion}) directly to the 
two-loop cross section calculated in full QCD. Both
ways of matching must be carried out for stable top quarks
($\Gamma_t=0$) and are performed in the (formal) limit
$\alpha_s\ll v\ll 1$ where NRQCD and full QCD are applicable in the
conventional multiloop approximation.\footnote{
Other matching points
like $v=0$ are also possible, see e.g.~\cite{Labelle2}).
}
In our case one has to match at the two-loop level including terms up
to NNLO in an expansion in $v$.
Technically the second way of matching, called ``direct
matching''~\cite{Hoang1}, is simpler because it allows for a sloppier
treatment 
of the regularization procedure. (In fact, using the first way one has
to be very careful to use exactly the same regularization for the
matching calculation as in the calculation of the correlators. This is
a quite tricky task if one wants to avoid solving the Schr\"odinger
equation in D dimensions.) The
disadvantage of direct matching, however, is that matching is carried
out at the level of the final result which practically eliminates the
possibility to use the calculated short-distance coefficient for any
other process. Further, it requires that a multiloop expression for
the cross section is at hand. For convenience, I use the ``direct
matching'' method.

\section*{Some Explicit Results}
\begin{figure}[t!] 
\begin{center}
\leavevmode
\epsfxsize=4cm
\epsffile[220 420 420 550]{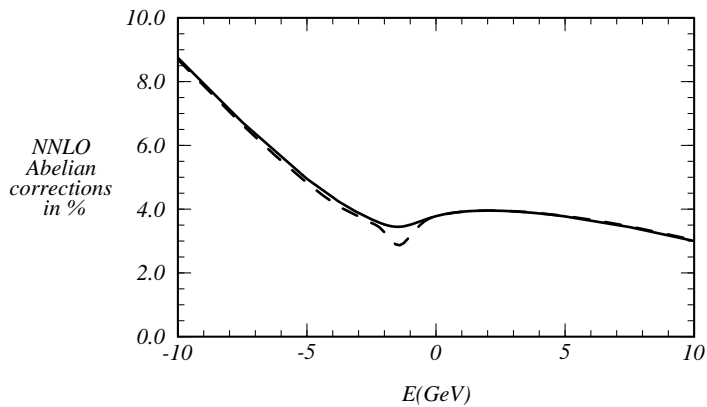}\\
%
%
\vskip  3.5cm
 \caption{\label{fig1} 
The NNLO Abelian corrections to the cross section in percent for
$\Gamma_t=1.56$~GeV (solid line) and $0.80$~GeV (dashed line). See the
text for more details.}
 \end{center}
\end{figure}

This talk would not be complete without making the somewhat general
discussion before more explicit. In the following I will carry out the
program described above for all Abelian contributions, i.e., for those
effects which also exist in QED. I will not present any technical
details and refer the interested reader to~\cite{Hoang2}, where the
calculations have actually been carried out.\footnote{
In Ref.~\cite{Hoang2} the calculations have not been formulated in the
framework of NRQCD. The results, of course, are not affected by this.
}
I start with the well-known expression for the Coulomb Green function
$G_c^{(0)}$~\cite{Wichmann1} 
\begin{eqnarray}
G_c^{(0)}(0,\vec r,\tilde E) & = &
-\,i\,\frac{M_t^2\,\tilde v}{2\,\pi}\,e^{i M_t \tilde v r}\,
\int_0^\infty dt \, e^{2 i M_t \tilde v r t}\,
\bigg(\frac{1+t}{t}\bigg)^{\frac{i a}{2 \tilde v}}
\,,
\quad
\tilde v\,\equiv\,\bigg(\frac{\tilde E}{M_t}\bigg)^{\frac{1}{2}}
\,,
\label{GreenfunctionLO}
\end{eqnarray}
The NNLO corrections coming from the kinetic term
$-\vec\nabla^4/4M_t^3$ and the Breit-Fermi potential $V_{\mbox{\tiny
BF}}$ are calculated via TIPT and read
\begin{eqnarray}
\Big[\, \delta G(0,0,\tilde E)
\,\Big]_{\mbox{\tiny Abel}}^{\mbox{\tiny NNLO}} & = &
\int d\vec x^3\, G_c^{(0)}(0,\vec x,\tilde E)\,
\Big[\,\frac{\vec\nabla^4}{4 M_t^3} - V_{\mbox{\tiny BF}}(\vec x)\,\Big]
\,G_c^{(0)}(\vec x,0,\tilde E)
\,.
\label{GreenfunctionBF}
\end{eqnarray}
Abelian corrections coming from the one- and two-loop
contributions in the Coulomb potential do not exist because we define
$\alpha_s \equiv \alpha_{s,\overline{\mbox{\tiny MS}}}^{(n_l=5)}$.
As mentioned before, taking the limit
$|\vec r|,|\vec r^\prime|\to 0$ in Eq.~(\ref{GreenfunctionLO}) and the 
integral~(\ref{GreenfunctionBF}) leads to UV divergences which I
regularize with the short-distance cutoff $\mu_{\rm fac}$. This
leads to 
the following result for the correlator ${\cal{A}}_1$ at NNLO
\begin{eqnarray}
\Big[{\cal{A}}_1\Big]_{\mbox{\tiny Abel}} & = &
\frac{3\,a\,M_t^2}{2\,\pi}\,
\bigg\{\,
i \,\tilde v - a \,\bigg[\,
\ln(-i \frac{M_t\,\tilde v}{\mu_{\rm fac}}) + \gamma 
  + \Psi\bigg( 1-i\,\frac{a}{2\,\tilde v} \bigg)
\,\bigg]
\,\bigg\}^2
\label{A1explicit}
\\[2mm] & &
+\frac{9\,M_t^2}{2\,\pi}\,
\bigg\{
i\,\Big(\tilde v+\frac{5}{8} \tilde v^3\Big)
 - a \Big(1+2\,\tilde v^2\Big) \bigg[
\ln(-i \frac{M_t\,\tilde v}{\mu_{\rm fac}}) + \gamma 
   + \Psi\bigg( 1-i\,\frac{a\,(1+\frac{11}{8}\,\tilde v^2)}
  {2\,\tilde v} \bigg)
\bigg]
\bigg\},
\nonumber
\end{eqnarray}
where $\gamma$ is the Euler constant and $\Psi$ the digamma function.
All power divergences $\propto \mu_{\rm fac}/M_t$ are freely dropped
and $\mu_{\rm fac}$ is defined in a way that
expression~(\ref{A1explicit})  
takes the simple form shown above. For the correlator ${\cal{A}}_2$
only the LO contribution in~(\ref{A1explicit}) is relevant and we
arrive at 
\begin{eqnarray}
{\cal{A}}_2 & = &
\tilde v^2\,\frac{9\,M_t^4}{2\,\pi}\,
\bigg\{\,
i\,\tilde v - a \,\bigg[\,
\ln(-i \frac{M_t\,\tilde v}{\mu_{\rm fac}}) + \gamma 
  + \Psi\bigg( 1-i\,\frac{a}{2\,\tilde v} \bigg)
\,\bigg]
\,\bigg\}
\,.
\label{A2explicit}
\end{eqnarray}
There are no non-Abelian contributions to ${\cal{A}}_2$. The Abelian
contributions to $C_1$ are calculated by expanding
expression~(\ref{crosssectionexpansion}), expanding (first) for small
$\alpha_s$ and (then) $v$, keeping terms up to order $\alpha_s^2$
and NNLO in the $v$ expansion,\footnote{
At this point we set $\mu_{\rm soft}=\mu_{\rm hard}$ because for
$\alpha_s\ll v\ll 1$ a distinction between the soft and the hard
scale is irrelevant.
}
and demanding equality to the corresponding two-loop cross section
calculated in full QCD~\cite{Hoang2,Hoang3} for
$\Gamma_t=0$ ($v=(E/M_t)^{1/2}$), 
\begin{eqnarray}
\Big[\,R_{\mbox{\tiny 2loop QCD}}^{\mbox{\tiny
NNLO}}\,\Big]_{\mbox{\tiny Abel}} & = &
N_c\,Q_t^2\,\bigg\{\,\bigg[\,
\frac{3}{2}\,v-\frac{17}{16}\,v^3
\,\bigg] +
\frac{C_F\,\alpha_s}{\pi}\,\bigg[\,
\frac{3\,\pi^2}{4}-6\,v+\frac{\pi^2}{2}\,v^2
\,\bigg]
\nonumber\\[2mm] & & 
+\, \alpha_s^2\,\bigg[\,
\frac{C_F^2\,\pi^2}{8\, v} - 3\,C_F^2 + 
\bigg(\, \frac{49\,C_F^2\,\pi^2}{192}  
   + \frac{3}{2}\,C_{\mbox{\tiny Abel}} - C_F^2\,\ln v
\,\bigg)\,v
\,\bigg]
\,\bigg\}
\,,
\label{Rtwoloop}
\end{eqnarray}
where 
$C_{\mbox{\tiny Abel}}= 
C_F^2 [ \frac{1}{\pi^2} (
\frac{39}{4}-\zeta_3 ) +
\frac{4}{3} \ln 2 - \frac{35}{18}] + 
C_F T [
\frac{4}{9} ( \frac{11}{\pi^2} - 1 ) ]$. 
The result for $C_1$ then reads
\begin{eqnarray}
\Big[\,C_1\,\Big]_{\mbox{\tiny Abel}} & = & 
1 - 4 C_F\frac{\alpha_s(\mu_{\rm hard})}{\pi} + 
\alpha_s^2(\mu_{\rm hard})\,
\bigg[\,
C_{\mbox{\tiny Abel}} 
+ \frac{2}{3}\,C_F^2\,\ln\Big(\frac{M_t}{\mu_{\rm fac}}\Big)
\,\bigg]
\,.
\label{C1explicit}
\end{eqnarray}
Expression~(\ref{C1explicit}) does not
contain any energy dependent (or even IR divergent) contributions
because NRQCD and QCD have the same low energy behavior, i.e., all the
energy dependence is contained in the correlators. Apart from the
dependence on the hard scale $\mu_{\rm hard}$, which remains because 
$[C_1]_{\mbox{\tiny Abel}}$ represents a truncated perturbative
series,\footnote{
The dependence on $\mu_{\rm hard}$ of the ${\cal{O}}(\alpha_s)$ term 
in Eq.~(\ref{C1explicit}) is not cancelled by terms in the
${\cal{O}}(\alpha_s^2)$ contributions because non-Abelian and massless
quark corrections are not considered.
}
there is a dependence on the factorization scale $\mu_{\rm fac}$. As can be
seen in Eq.~(\ref{A1explicit}), for $\Gamma_t=0$ this dependence is
cancelled by a corresponding $\mu_{\rm fac}$-dependent term in
$[{\cal{A}}_1]_{\mbox{\tiny Abel}}$.\footnote{
The invariance under changing the factorization
scale $\mu_{\rm fac}$ can be used to resum renormalization group logarithms.
However, I would like to warn the reader to blindly apply
renormalization group methods in the belief this would represent a
resummation of ``leading logarithms''. Although it is true that a naive
resummation of logarithms is possible in this way, the resummed
logarithms would certainly not be ``leading''. This is a consequence
of the fact that $Q\bar Q$ systems in the threshold regime are
multi-scale problems. At the NNLO level, where all interactions can be
treated as instantaneous, only the relative momentum of the top quarks
$\sim M_t\alpha_s$ and the top quark mass $M_t$ are relevant
scales. At NNNLO, however, also the energy of the top quarks $\sim
M_t\alpha_s^2$ arises as a relevant scale and leads to a much more
complicated structure of the anomalous dimensions. It is in fact not
clear whether not even lower scales $~M_t\alpha_s^n$ ($n>2$) become
relevant if effects beyond the NNLO level are considered.
}
For $\Gamma_t\ne 0$ there is a small contribution 
$\propto a\frac{\Gamma_t}{M_t}\ln\mbox{$\frac{M_t}{\mu_{\rm fac}}$}$
which is not cancelled. This ambiguity arises from our ignorance of a
consistent treatment of the NNLO finite width effect mentioned at the
beginning.
In Fig.\ \ref{fig1} the relative size (in $\%$) of the NNLO Abelian
contributions  
$(R_{\mbox{\tiny NNLO}}^{\mbox{\tiny thr}}-
R_{\mbox{\tiny NLO}}^{\mbox{\tiny thr}})/
R_{\mbox{\tiny NLO}}^{\mbox{\tiny thr}}$
($R_{\mbox{\tiny NLO}}^{\mbox{\tiny thr}}$ contains only the
contribution from the Coulomb Green function,
Eq.~(\ref{GreenfunctionLO}), in ${\cal{A}}_1$ and the terms in $C_1$
up to ${\cal{O}}(\alpha_s)$) is
plotted in the energy range $-10$~GeV$\,< E < 10$~GeV for
$M_t=175$~GeV, $\alpha_s(M_z)=0.118$ and $\Gamma_t=1.56$~GeV (solid
line)/$0.80$~GeV (dashed line). For the scales the choices 
$\mu_{\rm soft} = 30$~GeV and $\mu_{\rm hard} = \mu_{\rm fac} = M_t$ have been
made and two-loop running of the strong coupling has been used. It is
evident that the corrections are indeed at the level of several
percent and that they are fairly insensitive to the value of
$\Gamma_t$ indicating that the size of the Abelian NNLO contributions
is not affected by the ignorance of a consistent treatment of the finite
width effects. I also would like to note that the largest source of
theoretical uncertainty in the cross section 
$R_{\mbox{\tiny NNLO}}^{\mbox{\tiny thr}}$ comes from the dependence
on the soft scale $\mu_{\rm soft}$ -- clearly because it governs the
perturbative series describing the long-distance (i.e.\ low energy)
effects for which the convergence can be expected to be worse
than for the short-distance coefficients. A thorough examination of
this behavior, however, has to wait until all the NNLO relativistic
corrections are calculated.
\section*{Conclusion}
Due to the large top quark mass the $t\bar t$ system in the threshold
regime offers a unique opportunity to study strong interactions in
heavy-quark--antiquark pairs in the threshold regime using
perturbative QCD. In this 
talk I have shown that NRQCD provides an ideal framework to determine
the $t\bar t$ cross section at threshold at future lepton
colliders. NRQCD, an effective field theory of QCD, allows for the
calculation of the cross section (including various distributions) to
in principle arbitrary precision by offering a systematic formalism
which parameterizes all higher order effects from first principles
QCD. In this respect NRQCD is superior to the present day
potentialmodel-like approach used for analyses of $t\bar t$ production
at threshold because NRQCD does not necessarily rely on any
phenomenological 
input. I therefore propose that the potentialmodel-like approach
should eventually be abandoned. In this talk I have given a detailed
recipe how NNLO relativistic corrections to the total vector current
induced cross section can be calculated using NRQCD, and I have
presented explicit results for all Abelian NNLO contributions.\\[2mm]
This work is supported in part by the U.S.~Department of Energy under
contract No.~DOE~DE-FG03-90ER40546.

\sloppy
\raggedright
\def\app#1#2#3{{\it Act. Phys. Pol. }{\bf B #1} (#2) #3}
\def\apa#1#2#3{{\it Act. Phys. Austr.}{\bf #1} (#2) #3}
\def\lhc{Proc. LHC Workshop, CERN 90-10}
\def\npb#1#2#3{{\it Nucl. Phys. }{\bf B #1} (#2) #3}
\def\nP#1#2#3{{\it Nucl. Phys. }{\bf #1} (#2) #3}
\def\plb#1#2#3{{\it Phys. Lett. }{\bf B #1} (#2) #3}
\def\prd#1#2#3{{\it Phys. Rev. }{\bf D #1} (#2) #3}
\def\pra#1#2#3{{\it Phys. Rev. }{\bf A #1} (#2) #3}
\def\pR#1#2#3{{\it Phys. Rev. }{\bf #1} (#2) #3}
\def\prl#1#2#3{{\it Phys. Rev. Lett. }{\bf #1} (#2) #3}
\def\prc#1#2#3{{\it Phys. Reports }{\bf #1} (#2) #3}
\def\cpc#1#2#3{{\it Comp. Phys. Commun. }{\bf #1} (#2) #3}
\def\nim#1#2#3{{\it Nucl. Inst. Meth. }{\bf #1} (#2) #3}
\def\pr#1#2#3{{\it Phys. Reports }{\bf #1} (#2) #3}
\def\sovnp#1#2#3{{\it Sov. J. Nucl. Phys. }{\bf #1} (#2) #3}
\def\sovpJ#1#2#3{{\it Sov. Phys. LETP Lett. }{\bf #1} (#2) #3}
\def\jl#1#2#3{{\it JETP Lett. }{\bf #1} (#2) #3}
\def\jet#1#2#3{{\it JETP Lett. }{\bf #1} (#2) #3}
\def\zpc#1#2#3{{\it Z. Phys. }{\bf C #1} (#2) #3}
\def\ptp#1#2#3{{\it Prog.~Theor.~Phys.~}{\bf #1} (#2) #3}
\def\nca#1#2#3{{\it Nuovo~Cim.~}{\bf #1A} (#2) #3}
\def\ap#1#2#3{{\it Ann. Phys. }{\bf #1} (#2) #3}
\def\hpa#1#2#3{{\it Helv. Phys. Acta }{\bf #1} (#2) #3}
\def\ijmpA#1#2#3{{\it Int. J. Mod. Phys. }{\bf A #1} (#2) #3}
\def\ZETF#1#2#3{{\it Zh. Eksp. Teor. Fiz. }{\bf #1} (#2) #3}
\def\jmp#1#2#3{{\it J. Math. Phys. }{\bf #1} (#2) #3}
\def\yf#1#2#3{{\it Yad. Fiz. }{\bf #1} (#2) #3}

\end{document}